\documentclass[%
 reprint,
 amsmath,amssymb,
 aps,
]{revtex4-1}

\usepackage{graphicx}% Include figure files
\usepackage{dcolumn}% Align table columns on decimal point
\usepackage{bm}% bold mat
\usepackage{color}
\usepackage{graphicx}
\begin{document}

\title{The effect of Ta oxygen scavenger layer on HfO$_2$-based resistive switching behavior: thermodynamic stability, electronic structure, and low-bias transport} %\\%Article title goes here instead of the text "This is the title"
%\vspace{0.3cm} & \vspace{0.3cm} \\

 %& 
\author{X. Zhong}
\affiliation{Materials Science Division, Argonne National Laboratory, Lemont, Illinois 60439, USA
}%
 
\author{I. Rungger}
\affiliation{
{Materials Division, National Physical Laboratory, Teddington, TW11 0LW, United Kingdom}
}%

\author{P. Zapol}
\affiliation{
 Materials Science Division, Argonne National Laboratory, Lemont, Illinois 60439, USA
}%

\author{H. Nakamura}
\affiliation{
 Nanosystem Research Institute (NRI), “RICS”, National Institute of Advanced Industrial Science and Technology
(AIST), Central 2, Umezono 1-1-1, Tsukuba, Ibaraki, 305-8568, Japan
}%

\author{Y. Asai}
\affiliation{
 Nanosystem Research Institute (NRI), “RICS”, National Institute of Advanced Industrial Science and Technology
(AIST), Central 2, Umezono 1-1-1, Tsukuba, Ibaraki, 305-8568, Japan
}%

\author{O. Heinonen}
 \email{heinonen@anl.gov}
\affiliation{
 Materials Science Division, Argonne National Laboratory, Lemont, Illinois 60439, USA
}%
\affiliation{
 Northwestern-Argonne Institute for Science and Engineering, Northwestern University, 2145 Sheridan Rd., Evanston, Illinois 60208, USA
}%

%\includegraphics{head_foot/dates} 
%& 
\begin{abstract}
Reversible resistive switching between high-resistance and low-resistance states in metal-oxide-metal heterostructures makes them very interesting for applications in random access memories. While recent experimental work has shown that inserting a metallic "oxygen scavenger layer" between the positive electrode and oxide improves device performance, the fundamental understanding of how the scavenger layer modifies heterostructure properties is lacking.  We use density functional theory to calculate thermodynamic properties and conductance of TiN/HfO$_2$/TiN heterostructures with and without Ta scavenger layer.  First, we show that Ta insertion lowers the formation energy of low-resistance states. Second, while the Ta scavenger layer reduces the Schottky barrier height in the high-resistance state by modifying the interface charge at the oxide-electrode interface, the heterostructure maintains a high resistance ratio between high- and low-resistance states. Finally, we show that  the low-bias conductance of device on-states becomes much less sensitive to the spatial distribution of oxygen removed from the HfO$_2$ in the presence of the Ta layer. By providing fundamental understanding of the observed improvements with scavenger layers, we open a path to engineer interfaces with oxygen scavenger layers to control and enhance device performance. In turn, this may enable the realization of a non-volatile low-power memory technology with concomitant reduction in energy consumption by consumer electronics and significant benefits to society.
%Metal-oxide-metal heterostructures exhibit reversible resistive switching between high-resistance and low-resistance states. This makes such structures very interesting for applications in random access memories. Recent experimental work has shown that inserting a metallic ``oxygen scavenger layer'' between the top (positive) TiN electrode and HfO$_2$ affects the conduction properties and improves device performance. We show, using atomistic modeling within 
%%the GGA+U scheme of 
%Density Functional Theory, that a Ta oxygen scavenger layer significantly enhances the thermodynamic stability of depleting oxygen from the oxide by lowering the formation energy of low-resistance states. The Ta scavenger layer also reduces the Schottky barrier height in the insulating system by modifying the interface charge at the oxide-electrode interface but maintains a high resistance ratio between high- and low-resistance states. Finally, the low-bias conductance of device on-states is insensitive to the location of oxygen removed from the HfO$_2$. Our work provides fundamental understanding of the experimentally observed improvements with scavenger layers, and also suggests more broadly how engineered interfaces with oxygen scavenger layers can provide a path to control and enhance device performance as scavenger layers may improve switching behavior and stability of high- and low-resistance states without degrading the resistance ratio between them. Therefore, oxygen scavenger layers may enable the realization of a non-volatile low-power memory technology with concomitant reduction in 
energy consumption by consumer electronics and significant benefits to society.
% \\%The abstrast goes here instead of the text "The abstract should be..."

%\end{tabular}

\end{abstract}

\maketitle

%%%FOOTNOTES%%%

%%%END OF FOOTNOTES%%%

%%%MAIN TEXT%%%%
%The main text of the article\cite{Mena2000} should appear here.

\section{Introduction}
Memristive switching devices are candidates for the next-generation fast, scalable, non-volatile, low-power memories \cite{yang2013memristive}. A typical device structure consists of an insulating metal oxide layer sandwiched between two electrodes, forming a metal-insulator-metal (MIM) heterostructure. While the as-deposited device is usually insulating, a low resistance state (the on-state) can be achieved after an ``electroforming" process \cite{Yang2009}. This is essentially a controlled soft dielectric breakdown during which conducting pathways are formed under the application of an electric field.  The device can subsequently be switched reversibly between the low-resistance on-state and a high-resistance off-state using voltage or current pulses. In most devices that depend on the movement of anions, oxygen has a relatively high mobility\cite{yang2013memristive} and the switching is understood to be caused by the migrating of oxygen from oxide to electrodes. In this work, we will focus on elucidating electrode-oxide interface properties in both low and high resistance states.

Many oxide-electrode materials combinations exhibit memristive switching, most of which do not satisfy requirements for implementation in commercial devices. 
%These include, among others, endurance, repeatability, high on-off resistance ratio (the ratio of the resistance in the off-state to that of the on-state), and good process control. 
The TiN-HfO$_2$ combination is a promising one because of its high scalability, and also because of very mature fabrication technologies based on decades of experience in processing both TiN and HfO$_2$ in the semiconductor industry. Achieving a robust and high (typically several orders of magnitude) on-off ratio (the ratio of the resistance in the high-resistance off-state to that of the low-resistance on-state) is essential for  reliable commercial memories, but in TiN-HfO$_2$-TiN heterostructures the on-state conductance is low and is characterized as semiconducting, {\em i.e.,} the conductivity is highly temperature-dependent\cite{Francesca2013}. 

Recently, it has been observed that the memristive switching properties of the TiN-HfO$_2$-TiN structure can be greatly improved by inserting an ``oxygen scavenger'' metal layer ({\em e.g.,} Hf) between TiN and HfO$_2$ \cite{Francesca2013, Simoen2013}. The on-state current has been shown to be enhanced by three orders of magnitude with little temperature dependence. It has been proposed that oxygen scavenger layers help increase the oxygen vacancy (Vo) concentration inside HfO$_2$, which in turn facilitates conducting filament formation \cite{Francesca2013}. The observed improved conductivity by inserting an oxygen scavenger layer strongly suggests the importance of rational design of electrodes for better device performance\cite{NakamuraIEEE2014}. 
%In fact, recent experiments on several oxide systems have revealed that the switching properties of an anion device is a property of the whole electrode-oxide system, rather than a property only of the oxide itself\cite{yang2011metal,peng2013effects, stille2012detection, chen2013migration, wang2014conducting, liu2010controllable}. In general, better conductivity in the on-states and smaller dispersion of on/off resistance ratio have been observed when the oxide is easily reduced by the contacting metal electrodes\cite{yang2011metal,peng2013effects, stille2012detection}, and/or when diffused oxygen is easily absorbed \cite{chen2013migration}. %  

In order to control device performance it is essential to understand in detail the device structure-properties relationship. However, characterization of the whole MIM device with atomic resolution is in general difficult \cite{Yajima2010} and has not been performed in previous works on HfO$_2$-TiN \cite{Francesca2013, Simoen2013, Xavier2012, Xavier2013}. In this work, we perform first-principles modeling to elucidate the effects of a thin tantalum layer inserted between the TiN cathode and the HfO$_2$ oxide layer in TiN-HfO$_2$ heterostructures. Tantalum is selected because it is also an oxygen scavenger with properties similar to hafnium, and has recently been shown to exhibit stronger electron coupling with HfO$_2$\cite{Nakamura2014}. The addressed device properties include thermodynamic stability, local chemical composition, electronic structure and transport. The thermodynamic stability of the high- and low-resistance states strongly impacts the device reliability and also strongly influences the local chemical structure of the device. The local chemical structure, in turn, directly determines the electronic structure and transport properties. We first study the thermodynamic stability of several prototypical atomic configurations of both TiN-HfO$_2$-TiN and TiN-Ta-HfO$_2$-TiN structures to establish a correlation between atom arrangement and stability of on- and off-states. In particular, we are interested in how oxygen is distributed at the TiN electrode-HfO$_2$ interface and, possibly, inside the scavenger layer or electrode after oxygen is moved out of HfO$_2$. Next, we examine the energy band offsets for the modeled MIM structures. The energy band offsets usually profoundly affect the electronic transport properties of the device, especially in the high-resistance off-state. We study how the energy levels of the oxide are shifted when Ta is inserted, and also how they are affected by the location of the diffused oxygen. Finally, we directly correlate the atomic models with the electronic transport properties of the device using a filamentary model for the
conductive on-state \cite{Kenji2012,Nakamura2014,NakamuraIEEE2014}. Our work shows that the oxygen scavenger layer reduces the formation energy of the conducting state and therefore improves its thermodynamic stability. Furthermore, our results show that the low-resistance state in the presence of a scavenger layer is metallic, which improves the on-off ratio and makes
the device performance more robust with much less dependence on temperature, without much
dependence on where the oxygen atoms are located in the low-resistance state. These improvements by the insertion of a scavenger layer are important enablers for the realization of non-volatile memory technologies as they may reduce voltage and power requirements for formation and operation, and may improve resistance distributions in low- and high-resistance states. 
Memory technologies based on resistive switching are interesting not only because their potential scalablility and non-volatility. The reduction in electronics power consumption by the introduction of non-volatile memories also has huge benefits for society as a whole. Our work puts the use of scavenger layers as an enabler for resistive switching memory technologies on a firm scientific footing.

\section{METHODS}

Calculations in the present work are based on Density Functional Theory (DFT) \cite{Kohn1964, Sham1965} using the Generalized Gradient Approximation (GGA) \cite{Wang1996} together with on-site $U$ parameters \cite{Sawatzky1994}, referred to as GGA+U, as implemented in the SIESTA/Smeagol electronic structure and transport codes \cite{Soler2002, Ivan2008, rocha2005, rocha2006, Pemmaraju2007}. GGA+U and other advanced DFT methods such as hybrid functionals and Self-Interaction Corrections (SIC) systematically alleviate the problems of underestimating semiconductor band gaps \cite{Zhong2015, Pemmaraju2007, Blanka2010, Broqvist2006}. We use norm-conserving pseudopotentials \cite{Troullier} and a plane wave cut-off at 400 Ry; a $2\times2\times1$ $k$-point sampling is used for geometry relaxation and transport calculations. To speed up the calculation of these relatively large models, we adopt a single zeta (SZ) basis set for Ti, O and N, and an SZ plus polarization (SZP) basis set for Ta and Hf. We have confirmed that the adopted basis sets yield a good description of the materials geometry (Table~\ref{table:length}), with the geometry relaxed to residual atomic forces smaller than 0.05~eV/{\AA}.

\begin{table}[h!]
\caption{A comparison of lattice constants in {\AA} calculated using the adopted basis set (SZ and SZP, see text), the default high quality DZP basis set, and measured experimentally.}
\begin{center}
\begin{tabular}{ | c | c | c | c | }
 \hline
  &\hspace{0.2cm}TiN\hspace{0.2cm} &\hspace{0.2cm}cHfO$_2$\hspace{0.2cm}&\hspace{0.2cm}Ta\hspace{0.2cm} \\
 \hline
  \hspace{0.2cm}Adopted basis\hspace{0.2cm}&4.32&5.10&3.36\\
 \hline
 DZP basis&4.27&5.08&3.32\\
 \hline
 Experimental&4.25\cite{Crecelius1984}&5.12\cite{passerini1930}&3.30\cite{waseda1975}\\
 \hline
\end{tabular}
\end{center}
\label{table:length}
\end{table}

In order to minimize interface strain of the repeating unit cell (see Fig.~\ref{fig:structure}), the (110) direction of cubic HfO$_2$ (cHfO$_2$), the (100) direction of rock salt TiN, and the (100) direction of Ta in the body-centered cubic phase are aligned with the vertical direction of the device (the electron transport direction), taken to be the $\hat z$-axis. The structure includes 550 atoms: 169 Ti, 169 N, 54 Hf, 108 O, and 50 Ta in one computational unit cell. In the xy-plane, the lattice constants of this unit cell are 10.81~{\AA} along the $\hat x$-axis, and 10.20~{\AA} along $\hat y$-axis, respectively, corresponding to $3 \times 2$ unstrained (110) cHfO$_2$ unit cells. With HfO$_2$ set at its experimental lattice constants, TiN is under a strain of +0.1\% and -5.6\% in the {$\hat x$}- and {$\hat y$}-directions, while Ta is under a strain of +3.7\% and -2.3\% in the $\hat x$- and $\hat y$-directions, respectively\cite{Crecelius1984, passerini1930, waseda1975}. Previous calculations\cite{Young2013} indicate that a 5\% strain only marginally changes the electrode work function (less than a 0.05 eV change). As a result, the induced strain in the present model because of lattice mismatch is not expected to significantly change the energy band alignment at the interfaces. For transport calculations, we attach semi-infinite TiN to both ends of the computational unit cell. 

In addition to requiring a good description of the material geometry, we use three additional criteria to identify optimal U-parameters, 8 (3) eV for the Hf 5d (O 2p) orbitals: (i) The calculated band gap for pristine cHfO$_2$ is 5.4~eV, in close agreement with the value of 5.2~eV obtained using a much more computationally expensive GW approximation \cite{Jiang2010}; (ii) The calculated energy level of a neutral single oxygen vacancy in monoclinic HfO$_2$ (mHfO$_2$) agrees well with previous experimental and theoretical work \cite{Gavartin2006, Broqvist2006}; we also predict the neutral vacancy level to be located just below the center of band gap. (We use oxygen vacancies in mHfO$_2$ for comparison and calibration because, to the best of our knowledge, there are no data available for vacancy levels in cHfO$_2$.)  This is important because we use a linear chain formed by neutral oxygen atoms as a model for a conducting filament, similar to the previous work\cite{Nakamura2014, Kenji2012}; and (iii) we obtain a Schottky Barrier Height (SBH) of 2.4~eV for the TiN-HfO2 interface, in good agreement with reported values  of 1.8 to 2.5~eV from experiments\cite{Cimino2012, Fonseca2006}.

\section{Results and discussion}

\subsection{Thermodynamics and the effect of the oxygen scavanger layer}

\begin{figure}[!ht]
\centering
\includegraphics[width=0.35\textwidth]{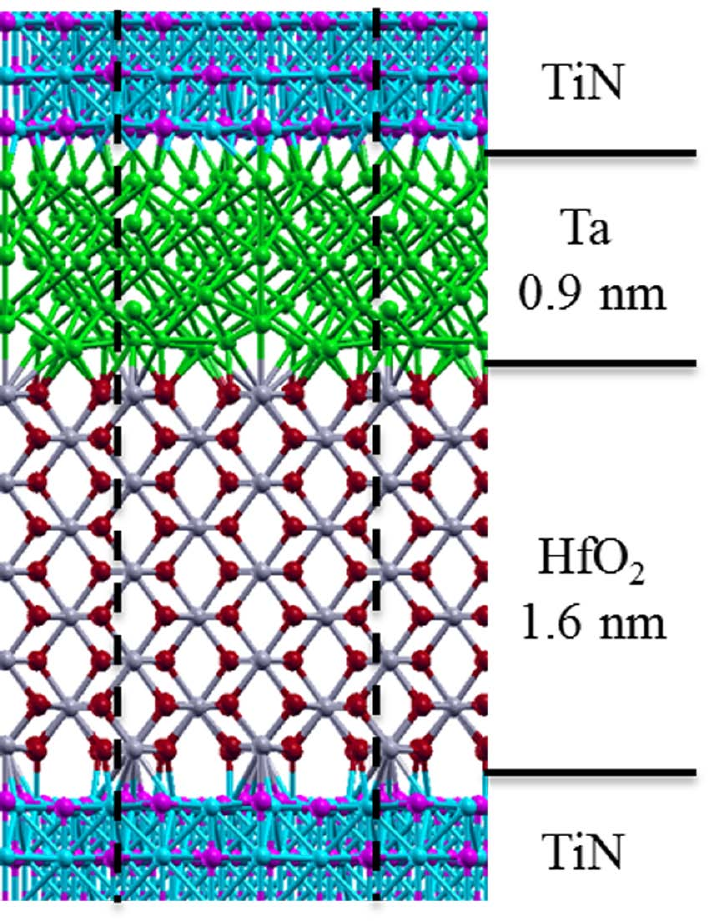}%{structure}
\caption{The relaxed atomic structure of the virgin TiN-HfO$_2$ device with tantalum layer insertion. The dashed lines show one unit cell used in the modeling. Color scheme: Ti in light blue, N in purple, Ta in green, Hf in grey and O in red.}
\label{fig:structure}
\end{figure}

The modeled virgin TiN-HfO$_2$ heterostructure with a Ta
insertion layer is shown in Figure~\ref{fig:structure}. The virgin structure is the reference structure with stoichiometric HfO$_2$ and no defects or oxygen vacancies. 

\begin{figure}[!ht]
\centering
\includegraphics[width=0.45\textwidth]{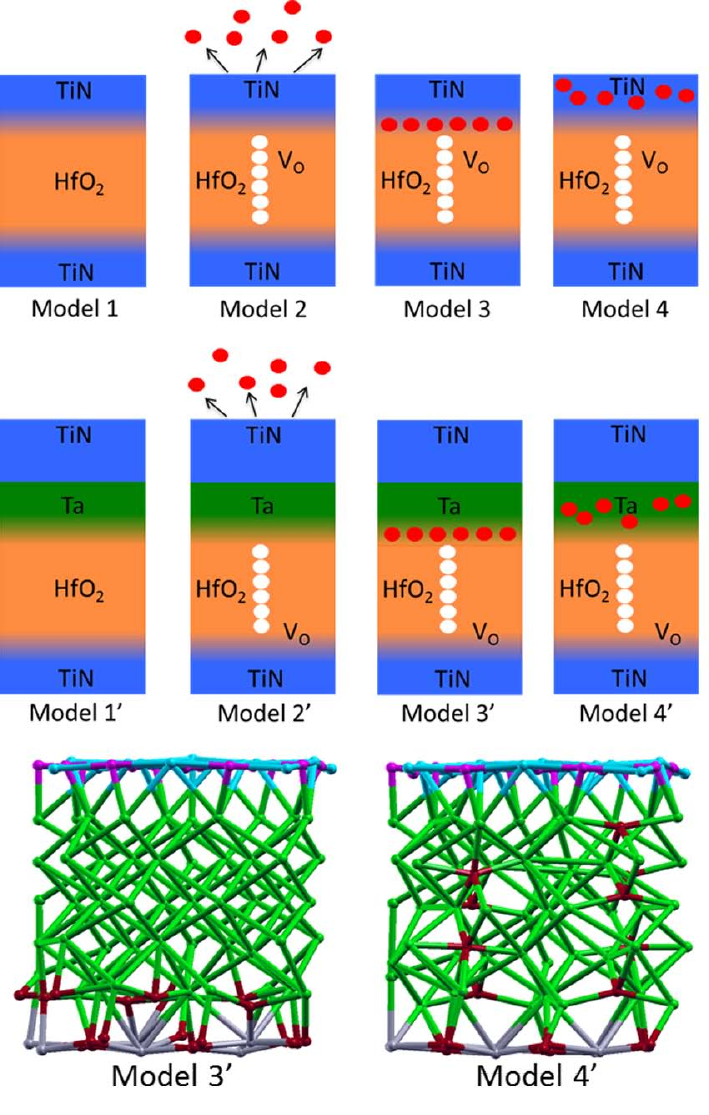}%{scheme}
\caption{Top and middle panels: cartoons of the models of memristor device. The top panels depict the TiN-HfO$_2$-TiN devices without Ta layer. From left to right: the virgin state with pristine HfO$_2$ (Model 1), oxygen vacancy filament formed in the HfO$_2$ matrix with oxygen escaping to ambient atmosphere (Model 2), filament formed with oxygen accumulating at the TiN-HfO$_2$ interface (Model 3) and, filament formed with oxygen distributing uniformly in TiN electrode (Model 4). The middle panel shows the same sequence of oxygen distribution arrangement but with Ta oxygen scavenger layer inserted (Models 1'-4'). The bottom panel shows the atomic structures of the interface region of Models 3' and 4', including one atomic layer of TiN, the Ta inserting layer and one atomic layer of HfO$_2$ (Ti in light blue, N in purple, Ta in green, Hf in grey and O in red).}
\label{fig:scheme}
\end{figure}

We have considered eight prototype states in two equal sets, the first set without a Ta scavenger layer and the second one with the layer (Fig.~\ref{fig:scheme}). We will refer to  either the TiN electrode in the first set or the TiN-Ta in the second set as ``electrode", and when we speak of the ``interface" between the oxide and the electrode, we refer to the oxide-TiN interface in the first set, and the oxide-Ta interface in the second set. Each set of states contains a virgin state with stoichiometric pristine cHfO$_2$ representing the off-state as a reference. Three atomic configurations within each set are used to represent the on-state. In these on-states a linear oxygen vacancy filament is simulated by removing one adjacent oxygen atom from each of the nine HfO$_2$ layers. This linear filament model of oxygen vacancy is similar to that adopted for monoclinic HfO$_2$ (mHfO$_2$) to study the quantized conductance observed for HfO$_2$ \cite{Xavier2013}. The nine removed oxygen atoms are either simply taken out of device, simulating the oxygen-escaped state (see Table~\ref{table:stability}), or placed at the interface between electrode and HfO$_2$ (``O at interface" in  Table~\ref{table:stability}), or distributed inside the electrode (``O inside electrode" in Table~\ref{table:stability}). These three cases are used to represent three ideal extremes of the device state. The energies per moved oxygen atom of the on-states relative to virgin TiN-cHfO$_2$-TiN and TiN-cHfO$_2$-Ta-TiN reference states are summarized in Table~\ref{table:stability}. For each device set (either with or without Ta scavenger layer), the relative energy of one device state ($S_i$ or $S_{i'}$, $i,i'=2,\ldots,4$) is calculated by the total energy difference between this state and the corresponding virgin state (\(S_1\)), i.e, \(\Delta E_{Si} =E_{Si}-E_{S1}\), where \(E_{Si}\) and \(E_{S1}\) are the total energies of the given device state and the virgin state, respectively (and analogously for the primed states). Note that for the states 2 and 2' for which oxygen escapes from device, the total energy is calculated by adding the total energy of the remaining device to the formation energy of the oxygen molecules. 

\begin{table}[h!]
\caption{Thermostatic stability: the energy difference (in eV) between different device states and reference virgin states. The energy difference is per oxygen atom for nine oxygen atoms removed from the oxide.}
\begin{center}
\begin{tabular}{ | c | c | c | }
 \hline
  &\hspace{0.2cm}TiN-HfO$_2$-TiN\hspace{0.2cm} &\hspace{0.2cm}TiN-Ta-HfO$_2$-TiN\hspace{0.2cm} \\
 \hline
  Virgin state (No Vo) &0.00 (Model 1) &0.00 (Model 1')\\
 \hline
% O escaped &8.93&8.97\\
 O escaped & 7.40 (Model 2) & 7.51 (Model 2')\\
 \hline
 O at interface&3.63 (Model 3) &2.09 (Model 3')\\
 \hline
 \hspace{0.2cm}O inside electrode\hspace{0.2cm} &8.93 (Model 4)&3.18 (Model 4')\\
 \hline
\end{tabular}
\end{center}
\label{table:stability}
\end{table}

When oxygen is placed at the TiN-HfO$_2$ interface, oxygen atoms move towards HfO$_2$ and form bonds with Hf during the geometry relaxation process, avoiding the TiN. In contrast, when oxygen atoms are placed at the Ta-HfO$_2$ interface they move to and bind with Ta (left bottom panel, Fig.~\ref{fig:scheme}). As oxygen has a very low solubility in TiN, we model the oxygen inside TiN (Model 4) by calculating the energy of placing one oxygen atom inside TiN bulk (modeled by a cubic box with side length of about 1.1 nm) and multiplying the total energy difference of the model with and without oxygen atom by nine. For the case where oxygen is distributed inside Ta, oxygen atoms are placed at the interstitial sites in the five Ta layers (right bottom panel, Fig.~\ref{fig:scheme}). Our results show that the virgin device state without a conducting filament is the most stable state thermodynamically, with or without Ta scavenger layer (Table~\ref{table:stability}). This implies that an appreciable energy needs to be supplied by electric field to move oxygen from their initial positions in HfO$_2$.

The hypothetical case of oxygen atoms escaping device and forming O$_2$ molecules is an endothermic process with an energy cost as high as 7.5~eV (per O atom) for the modeled systems, both with and without tantalum layer. The effect of the Ta oxygen scavenger layer is clear if oxygen is assumed to remain in the system without escaping to the atmosphere: when oxygen is moved from  HfO$_2$ to the Ta-HfO$_2$ interface, the energy cost is calculated to be as low as 2.1~ eV. This is less than two thirds of the corresponding energy cost of moving the oxygen to the TiN-HfO$_2$ interface (3.6~eV). This means a significantly lower cost in energy for removing oxygen from the HfO$_2$ matrix when the Ta oxygen scavenger layer is inserted. Interestingly, similar trend has been observed experimentally, {\em i.e.,} by inserting an Hf layer between TiN and HfO$_2$ the forming bias is reduced from 4 eV to about 2 eV \cite{Francesca2013}.

In contrast, the energy cost for moving oxygen into the TiN electrode is forbiddingly high (8.9~eV), which is even larger than that of oxygen escaping to the ambient atmosphere. Thus, once the interface is saturated with oxygen, further removal of oxygen from the HfO$_2$ is highly unfavorable, making the formation of an oxygen vacancy filament difficult in a HfO$_2$-TiN structure without scavenger layer. On the other hand, it is much easier for oxygen to move into Ta, with a much smaller energy cost (3.2~eV) than for moving into TiN. The lower energy cost for moving oxygen into the scavenger layer implies significantly improved switching properties in the presence of the scavenger layer by avoiding oxygen piling up at the electrode-HfO$_2$ interface, which restrains the formation of an oxygen-deficient phase in HfO$_2$. The inserted Ta oxygen scavenger layer thus works as a potential trap for oxygen, restraining oxygen from either escaping out of device or moving into the TiN electrode. As a result, oxygen is constrained to the active region near the electrode-oxide interface during device operation, potentially leading to a reduction of device variance and enhanced device endurance.    

\subsection{Energy band offsets}

We will now discuss the electronic structure of the MIM structures to elucidate the effect of the Ta insertion and location of diffused oxygen on the electronic structure. As we showed in the previous section, the on-states with Ta are favorable thermodynamically over the on-states without Ta, so we will therefore mostly focus on the on-states with Ta. We will study the virgin states both with and without Ta insertion layer (Models 1 and 1'), the lowest-energy on-state without Ta (Model 3), as well as the on-states with Ta (Models 2', 3' and 4'). We are specifically interested in how the energy levels of the sandwiched oxide are aligned with those of the metallic electrodes for the different proototypical systems.

\begin{figure}[!ht]
\centering
\includegraphics[width=0.5\textwidth]{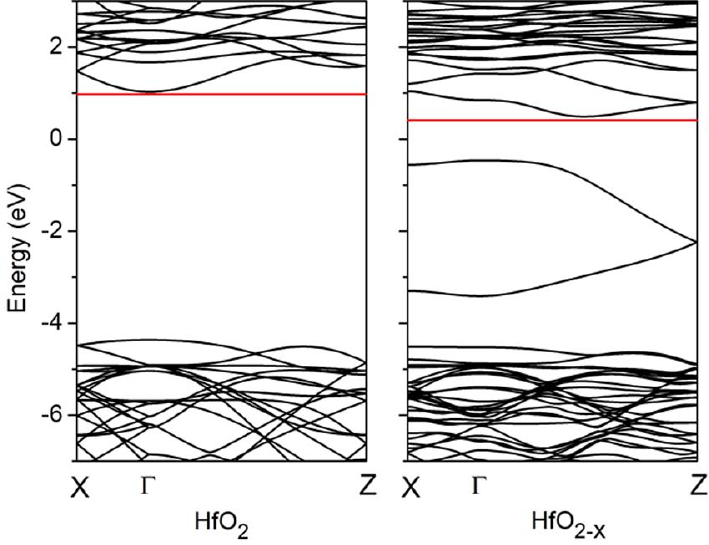}%{band}
\caption{Electronic band structure of cHfO$_2$ in the stoichiometric
phase (left panel), and with an oxygen vacancy
filament formed (right panel). The red lines denote the
conduction band minimum (CBM) of both phases.} 
\label{fig:band}
\end{figure}

Our model systems all have nine HfO$_2$ atomic layers with a total thickness of 1.6~nm sandwiched between electrodes in each model.
We first show the band structures we obtained for isolated HfO$_2$, both in the stoichiometric form and in the reduced form, in Fig.~\ref{fig:band}. Note that we will always assign zero of energy to the Fermi level. The electronic states near the Fermi level for the reduced HfO$_2$ are found to be dominated by vacancy-induced states, which form new mid-gap bands (right panel of Figure~\ref{fig:band}). As a result, the calculated energy gap is reduced from 5.4~eV to 0.8~eV, and there is a small density of states below the Fermi level but above the valence bands. 

\begin{figure}[!ht]
\centering
\includegraphics[width=0.45\textwidth]{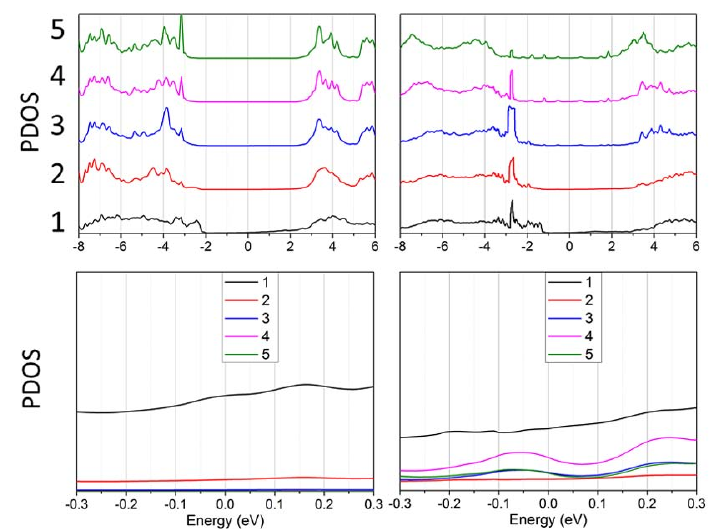}%{pdos1}
\caption{Projected density of states (PDOS) for Model 1 (left) and Model 3 (right) in arbitrary units. Curves 1 to 5 show the evolution of PDOS from the interface region to the oxide bulk region layer by layer, with layer 1 next to the electrode and layer 5 in the center of the oxide. Adjacent curves in the top graphs are offset vertically by 20 for clarity. The lower panels are the corresponding 20x enlargements around Fermi level (set to zero of energy.)}
\label{fig:pdos1}
\end{figure}

\begin{figure}[!ht]
\centering
\includegraphics[width=0.47\textwidth]{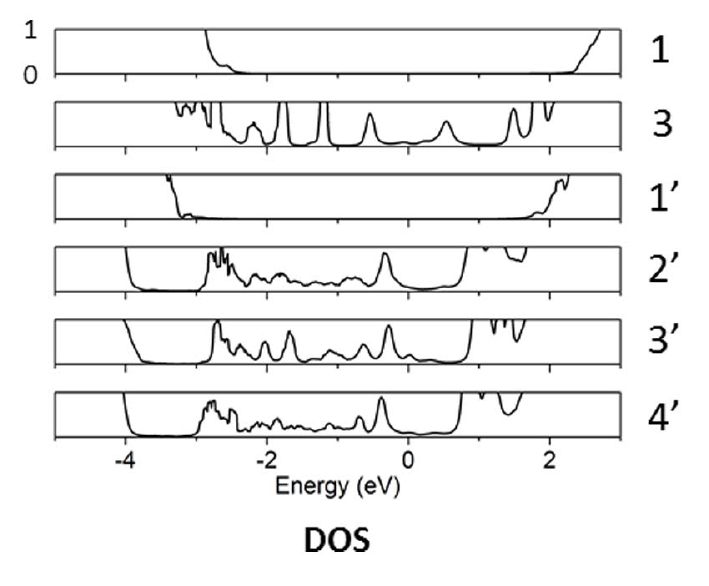}%{pdos}
\caption{Electronic structure of the oxide film sandwiched between metal electrodes for models 1, 3, 1' - 4', represented by the PDOS (arbitrary units) of the central oxide atomic layer. For all models we assign zero of energy to the Fermi level.}
\label{fig:pdos}
\end{figure}

We use the projected density of states (PDOS) of the central HfO$_2$ layer (layer 5) to investigate the electronic structure of the sandwiched HfO$_2$; in contrast with the central layer, the electronic states of the HfO$_2$ layers in direct contact with the electrodes hybridize with those of the electrodes, such that the local electronic structure and PDOS are strongly influenced by the electrodes. We show the evolution of the PDOS from the interface region to the central HfO$_2$ bulk region of Models 1 and 3 as examples in Fig.~\ref{fig:pdos1}. For the first two HfO$_2$ layers from the interface, layers 1 and 2, the local energy gap is closed even for Model 1, with an appreciable PDOS around the Fermi level (lower panels). From layer 3, the PDOS starts to converge to its values in the bulk region. In Models 2' - 4' each layer in the HfO$_{2-x}$ has a finite density of states (DOS) at the Fermi energy, E$_f$, indicating metallic properties across the reduced oxide film (not shown).  In addition, the contact of HfO$_{2-x}$ with Ta-TiN electrodes appears to induce small but finite PDOS throughout the energy range from the valence band maximum (VBM) to the conduction band minimum (CBM), which effectively closes energy gap as shown for the central layer PDOS in Fig.~\ref{fig:pdos}. This implies a weak metallic temperature dependence of the on-state and is consistent with the experimental observation that a scavenger layer reduces the temperature dependence of the on-state\cite{Francesca2013}.

We now address the energy band offset and Schottky barrier height (SBH) by analyzing the PDOS of the sandwiched oxide film (Fig.~\ref{fig:pdos}). The device Fermi level is set by the TiN electrodes, which are semi-infinite on both sides of the sandwiched oxide. For the \textit{virgin states} (Models 1 and 1') there is a clear gap of about 5 eV as in bulk HfO$_2$ (cf. Fig.~\ref{fig:band}, left panel), but the inserted Ta layer shifts the PDOS downward in energy (Model 1'). 
 The location of the CBM of the oxide with respect to the Fermi level gives the SBH. The SBH are 2.4 and 1.7~eV for Models 1 and 1', respectively, indicating a down-shift of 0.7~eV of the PDOS spectrum upon Ta insertion. For the on-state Model 3 without Ta, the CBM (and VBM) is shifted down in energy about 1~eV relative to the virgin state (Model 1), and there is a small but non-zero PDOS at the Fermi level $E_f$, in addition to isolated PDOS peaks below $E_f$. For the on-states with Ta (Models 2', 3', and 4'), the CBM is shifted down about 1.25~eV compared to the virgin state Model 1'. In addition there appear electron states induced in the gap region as a result of reduction of the HfO$_2$. The PDOS at E$_f$ is again finite as in Model 3, which indicates metallic properties. However, in contrast with Model 3, there is now a finite PDOS in the entire energy range from CBM (at about 0.75~eV above $E_f$) to about 3~eV below $E_f$. We also note that there are no significant differences in the PDOS spectra of Models 2', 3' and 4', with PDOS peaks at approximately the same energies for all these three models. This shows that the energy band offset is insensitive to the distribution of the diffused oxygen atoms in the presence of the Ta oxygen scavenger layer.

\begin{figure}[!ht]
\centering
\includegraphics[width=0.47\textwidth]{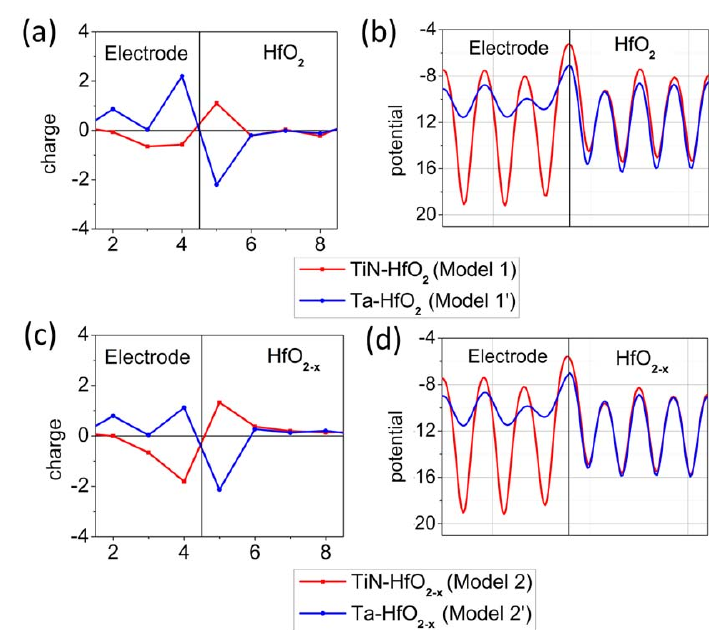}%{charge}
\caption{The charge distribution in electron per square nanometer of each atomic layer across electrode-oxide interface for Model 1 and 1' (a) and for Model 2 and 2' (c). Panel (b) represents the electrostatic (Hartree) potential energy (in eV) for Models 1 and 1', while panel (d) shows the electrostatic potential energy for Models 2 and 2'. For all panels three atomic layers of electrodes and four layers of HfO$_2$ are plotted. The vertical black lines denote the interface between electrode and HfO$_2$.}
\label{fig:charge}
\end{figure}

Our results are consistent with recent experimental work on the effect of inserting a hafnium layer between TiN and HfO$_2$ \cite{Jurczak2011}. In Ref.~\cite{Jurczak2011} it was observed that the SBH of the TiN-HfO$_2$ heterostructure was reduced by about 1~eV with the insertion of an Hf inter-layer, and attributed the SBH reduction to a surface dipole induced by oxygen scavenging by the inserted Hf layer. We will now show that our simulation strongly supports this conjecture. We plot the charge distribution across the electrode-HfO$_2$ interfaces of selected models in Fig.~\ref{fig:charge}. We begin our analysis by discussing Models 1 and 1' [panel (a) in Fig.~\ref{fig:charge}], which contain insulating stoichiometric HfO$_2$. The value of the SBH, or more generally, how the energy bands of the oxide are aligned with those of the electrode, is determined by the effective interface dipole \cite{tung2014physics}. Figure~\ref{fig:charge} (a) shows that without Ta insertion (Model 1), a negative charge appears on the electrode side, while there is a positive charge on the oxide side. As a result, an interface electric dipole is induced, pointing from electrode to oxide. In contrast, when Ta is inserted (Model 1'), there is a positive charge on the electrode (Ta) side and a negative charge on the oxide side, with the effective electric dipole pointing from oxide to electrode. This reflects the fact Ta has a strong tendency to lose electrons, because of its low electronegativity, compared to TiN. The \textit{net effect} of inserting Ta between TiN and HfO$_2$ is then a downward shift in energy of the average electrostatic (Hartree) potential energy in Model 1', compared to Model 1, as seen in Fig.~\ref{fig:charge} (b). This downward shift is about 0.7~eV, which is approximately equal to the downward shift in the SBH from Model 1 to Model 1'. Indeed, we obtain a decrease of 0.7~eV of the SBH of the virgin states [Models 1 and 1', Fig.~\ref{fig:pdos}] when Ta is inserted, which is comparable to the experimental value of 1 eV obtained for Hf-HfO$_2$ \cite{Jurczak2011}.  

For the on-states with reduced HfO$_2$, inserting Ta has a similar effect on the charge redistribution at the electrode-oxide interface [Fig.~\ref{fig:charge} (c)]. However, both for Models 2 and 2' the induced charge on the electrode side is reduced (smaller in magnitude for Model 2 and more negative for Model 2') compared to the charge in the virgin states Models 1 and 1'.
(We are here using Models 2 and 2' as they provide for a better comparison of what inserting Ta does than the other models with reduced oxide.)
However, the electrostatic potential energies rapidly become equal in the reduced oxide [Fig.~\ref{fig:charge} (d)]. 

\begin{figure}[!ht]
\centering
\includegraphics[width=0.35\textwidth]{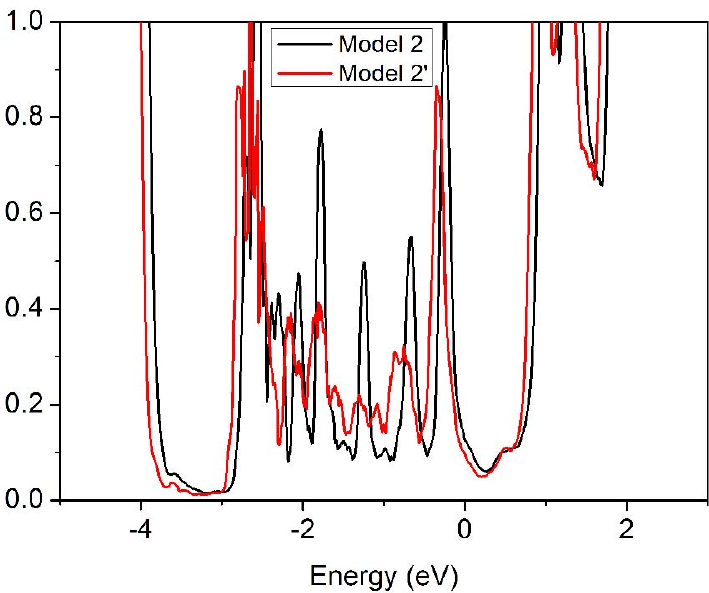}%{Vo-layer5}
\caption{PDOS of the central HfO$_{2-x}$ layer of Models 2 and 2': for the reduced HfO$_{2-x}$ phase, insertion of Ta has a negligible effect on the energy band offset.}
\label{fig:Vo-layer5}
\end{figure}

In contrast with stoichiometric HfO$_2$, reduced HfO$_{2-x}$ is metallic. When coupled with the electrodes, E$_f$ of HfO$_{2-x}$ and E$_f$ of electrodes must match each other to form the common Fermi level of the device. Thus, the PDOS for HfO$_{2-x}$ shifted in energy relative to that of HfO$_2$, with a magnitude of the shift determined by enforcing local charge neutrality. An electric field induced by an interface dipole is screened by free electrons in HfO$_{2-x}$. All modeled on-states with Ta show similar energy level alignment between oxide and electrodes (Models 2', 3' and 4',  Fig.~\ref{fig:pdos}), irrespective of where the diffused oxygen atoms are located. Furthermore, the Ta insertion has little effect on the PDOS for the reduced oxide when the oxygen is not at the oxide-electrode interface, as shown for Models 2 and 2' in Fig.~\ref{fig:Vo-layer5}. (Oxygen at the TiN-HfO$_{2-x}$ interface, Model 3, severely distorts the interface structure, which affects PDOS below $E_f$, as seen in Fig.~\ref{fig:pdos}.)

\subsection{Transport properties}

\begin{figure}[!ht]
\centering
\includegraphics[width=0.35\textwidth]{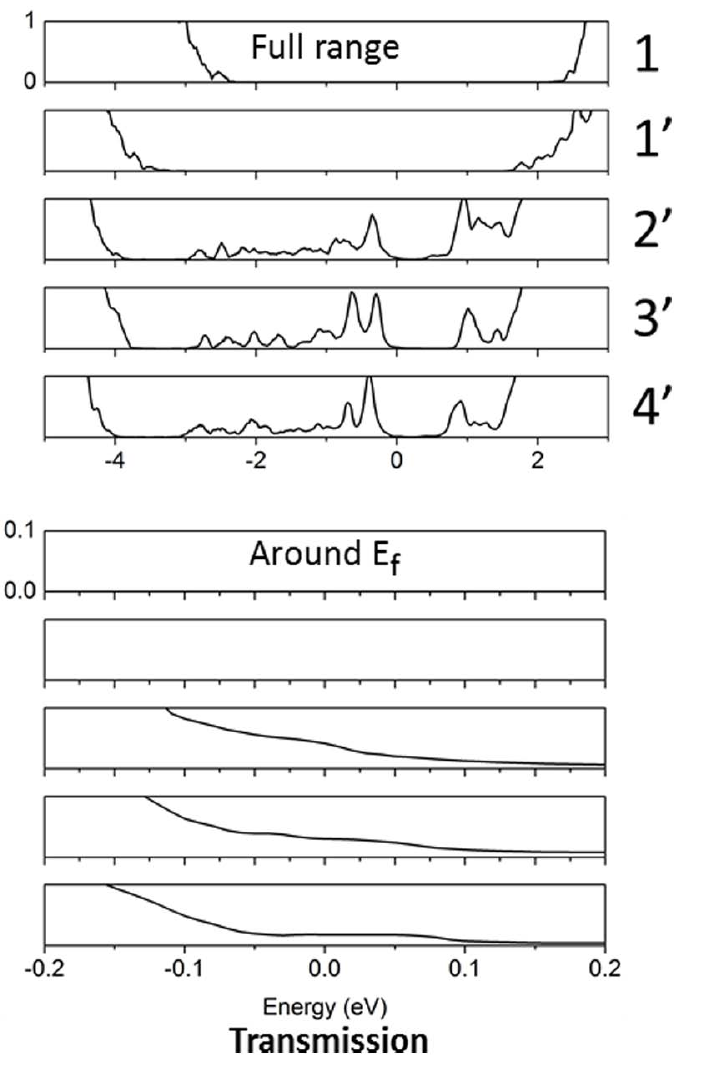}%{transport}
\caption{Electron transmission shown in both expanded energy range (upper panels, ``Full range"), and a close-up near $E_f$ (lower panels, ``Around E$_f$") in units of the conductance quantum. The zero of energy is set to $E_f$ in both panels.}
\label{fig:transport}
\end{figure}

Our calculated transmission spectra for the MIM structures (Fig.~\ref{fig:transport} ``Full range") show a clear correlation with the DOS of the sandwiched oxide film in Fig.~\ref{fig:pdos}. For the virgin states, the location of spectrum gap is very similar both in the transmission spectrum and in the PDOS spectrum. The values of the SBH estimated from the transmission spectra are nearly the same as those obtained from PDOS of the central oxide layer. Similarly, for the on-states major transmission peaks have similar locations as the peaks in PDOS across the different models.     

\begin{table}[h!]
\caption{The low-bias conductance of various models in unit of the conductance quantum ($G_0=2e^2/h$) per nm$^2$.}
\begin{center}
\begin{tabular}{ | c | c | c | c | c | c | }
 \hline
  Model&l&1'&2'&3'&4' \\
 \hline
 \hspace{0.cm}Conductance\hspace{0.cm} &\hspace{0.cm}3.7E-6\hspace{0.cm} &\hspace{0.cm}5.2E-6\hspace{0.cm} &\hspace{0.cm}6.3E-2\hspace{0.cm} &\hspace{0.cm}6.3E-2\hspace{0.cm} &\hspace{0.cm}3.6E-2\hspace{0.cm} \\
 \hline
\end{tabular}
\end{center}
\label{table:conductance}
\end{table}

\begin{figure}[!ht]
\centering
\includegraphics[width=0.45\textwidth]{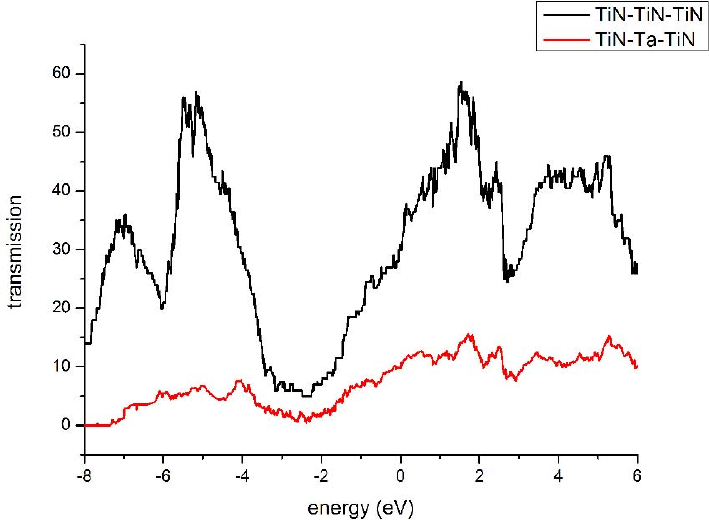}%{TiN-Ta-TiN}
\caption{Electron transmission function of a TiN-Ta-TiN structure (red) and of a reflection-less TiN-TiN-TiN structure (black). The transmission of the TiN-Ta-TiN structure is suppressed because of interfacial scattering because of band mismatch.}
\label{fig:TiN-Ta-TiN}
\end{figure}

We now discuss the low-bias conductance of selected models. We focus on transmission in an energy range from E$_f$-0.2~eV to E$_f$+0.2~eV (Fig.~\ref{fig:transport} ``Around E$_f$"). The virgin states show negligible transmission in this energy range, signaling transport only through quantum tunneling. The on-states in contrast show finite transmission at E$_f$, consistent with metallic properties. In order to quantitatively compare the  conductances of the different models, we list their low-bias conductances in Table~\ref{table:conductance}, in which the conductance is estimated by averaging the transmission in an energy range from E$_f$-0.2 to E$_f$+0.2~eV. For the virgin states, transport under low bias falls into the tunneling regime, with transmission at Fermi level on the order of $10^{-6}$ $G_0$ per nm$^2$. 
%Note that $G_0=2e^2/h$ corresponds to transport of a single ballistic conduction channel. 
Although the SBH is reduced by 0.7~eV when Ta is inserted, the effect on the low-bias transport is small: the estimated conductance of Models 1 and 1' is of the same order of magnitude. The tendency of enhanced tunneling current with reduced SBH when Ta is inserted is offset by electron scattering because of the band mismatch at the additional TiN-Ta interface. This is
illustrated in Fig.~\ref{fig:TiN-Ta-TiN}, which compares the transmission functions of two hypothetical structures, both with TiN electrodes. The TiN-TiN-TiN structure consists of a TiN part in the central scattering region, which of course couples seamlessly with two TiN electrodes without any interfacial scattering. As a result, electrons in the whole energy range can be transmitted in a ``reflectionless" way, i.e., without any scattering. 
Therefore, the transmission function for the TiN-TiN-TiN structure just reflects the DOS properties of TiN. In contrast, the mismatch of band structure between Ta and TiN results in a decreased transmission function for the TiN-Ta-TiN structure in the whole energy range compared to the reflectionless TiN-TiN-TiN structure.

\begin{figure}[!ht]
\centering
\includegraphics[width=0.45\textwidth]{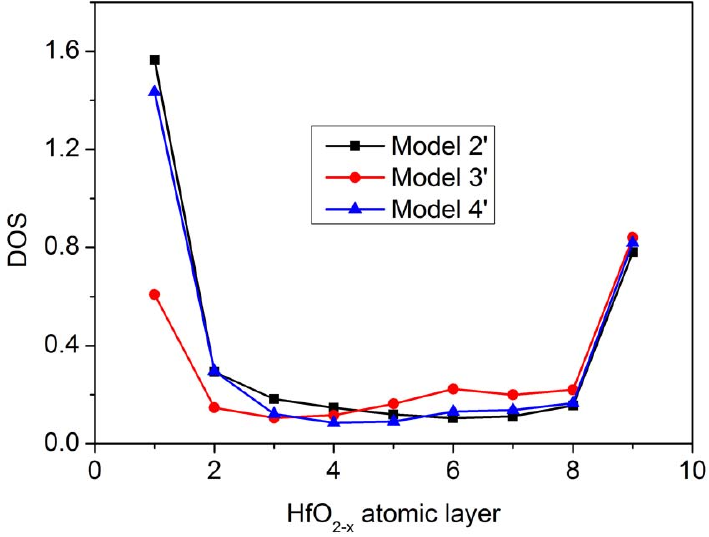}%{fluctuation}
\caption{Evolution of average local PDOS (arbitrary units) across HfO$_{2-x}$ atomic layers, averaged over the energy range between E$_f$+0.2~eV and E$_f$-0.2~eV}
\label{fig:fluctuation}
\end{figure}

For the on-states, all structures (Models 2' to 4') exhibit low-bias conductance on the order of $10^{-2}$ $G_0$ per nm$^2$, which is four orders of magnitude greater than that of the virgin states. Thus, the conductivity of the modeled on-states has no strong dependence on the location of diffused oxygen atoms. We note that diffused oxygen either inside Ta or at the Ta-oxide interface induces variations in the magnitude of available PDOS near the Fermi level across HfO$_{2-x}$ layers (Fig.~\ref{fig:fluctuation}). However, each atomic layer has a non-zero PDOS near $E_f$. As a result, the on-states are metallic, regardless of the location of the diffused oxygen. 

\section{CONCLUSION}

In summary, we have investigated the behavior of TiN-HfO$_2$-TiN and TiN-HfO$_2$-Ta-TiN heterostructures with a goal of elucidating the role of the Ta oxygen scavenger layer on thermodynamic stability, electronic structure and transport.
We find that the presence of the Ta layer improves the stability of the reduced oxide system, in which oxygen has been depleted from the HfO$_2$, and facilitates the formation of conducting channel in the oxide, consistent with experimental work\cite{Jurczak2011}.
%The reduced formation energy of the low-resistance states is consistent with experimental observations using Hf scavenger layers\cite{Jurczak2011}. 
Furthermore, the Ta layer reduces the Schottky barrier height for the insulating system because of a different interface charge at the Ta-HfO$_2$ interface than at the TiN-HfO$_2$ interface. Nevertheless, the reduced Schottky barrier height does not negatively impact the on-off ratio.
%reduce the calculated resistance in the off-state compared to the TiN-HfO$_2$ structure. \textcolor{red}{An important consequence is} that the on-off ratio is not degraded upon introduction of the Ta insertion layer.
%Furthermore, we find that oxygen vacancies introduced in bulk HfO$_2$ lead to a reduced band gap.
%due to the formation of bands of defect states in the energy gap. 
%Upon coupling to the metallic electrodes, the reduced Hf oxide becomes metallic with a finite DOS at the Fermi level. 
% (Moved to section 3.2) This implies a weak metallic temperature dependence of the on-state and is consistent with the experimental observation that a scavenger layer reduces the temperature dependence of the on-state.
Also, the electronic structure of the reduced oxide with a Ta layer becomes insensitive to the location of the oxygen atoms that are removed from the stochiometric oxide, with the consequence that the low-bias conductance in the on-state is insensitive to the location of the oxygen atoms. This is in contrast with the structure {\em without} a Ta layer, for which the electronic structure can vary quite significantly, depending on where the oxygen atoms are located. These results are consistent with experimental observations that a scavenger layer reduces the dispersion of on/off resistance ratio\cite{chen2013migration}.

Our results, which are based on detailed density functional-theory calculations using the GGA+U approximation, explain experimental results\cite{Nakamura2014} that demonstrate better device performance, in particular how the Ta layer facilitates the formation of conducting filaments, and how the Ta layer can lead to more robust devices, that is, less dependence on the microstructure of the device in its on-state. These results provide a firm foundation for further studies of oxygen scavenger layers of different thickness and chemical composition as enablers for realizing low-power resistive random access memory technologies. 

\begin{acknowledgements}
The work by X.Z. and O.H. was supported by U. S. DOE, Office of Science under Contract No. DE-AC02-06CH11357. P.Z. acknowledges support from the U.S. Department of Energy, Office of Science, Materials Sciences and Engineering Division. I.R. acknowledges financial support from the European Union's Horizon2020 research and innovation programme within the PETMEM project. We gratefully acknowledge the computing resources provided on Blues and Fusion, high-performance computing clusters operated by the Laboratory Computing Resource Center at Argonne National Laboratory. 
\end{acknowledgements}

%%%END OF MAIN TEXT%%%


%\balance


%%%REFERENCES%%%
\bibliography{rsc} %You need to replace "rsc" on this line with the name of your .bib file

\end{document}